\documentclass{ptapap}
\usepackage{amsmath}
\usepackage{amssymb}
\usepackage{color}

\usepackage{soul}

\author{Michal Zaja\v{c}ek}[CFT,MPIFR,Ph1]
\author{Andreas Eckart}[Ph1,MPIFR]
\author{Silke Britzen}[MPIFR]
\author{Bozena Czerny}[CFT]
\affil[CFT]{Center for Theoretical Physics, Polish Academy of Sciences, Al. Lotnikow 32/46, 02-668 Warsaw, Poland}
\affil[MPIFR]{Max-Planck-Institut für Radioastronomie (MPIfR), Auf dem Hügel 69, D-53121 Bonn, Germany}
\affil[Ph1]{I. Physikalisches Institut der Universität zu Köln, Zülpicher Strasse 77, D-50937 Köln, Germany}

\title{Distribution of Radio Spectral Slopes of Galaxies in Optical Diagnostic Diagrams}
\headtitle{Radio spectral slopes of galaxies}

\begin{document}

\maketitle

\begin{abstract}

For about 500 intermediate-redshift sources ($0.04<z<0.4$), whose radio flux densities at $1.4\,{\rm GHz}$ are larger than 10 mJy, we performed additional observations at 4.85 GHz and 10.45 GHz using 100-m Effelsberg telescope. Our radio-optical galaxies are located preferentially in the composite and AGN spectral classes in the narrow line optical diagnostic diagrams (ODD). In the analysis, we focused on the distribution of radio spectral indices of radio synchrotron power-law profiles, $S_{\nu}\propto \nu^{+\alpha}$, in the ODDs. Using different analysis techniques, both observationally motivated and machine-learning based, we found three distinct groups--clusters in the radio loudness, [OIII]/H$\beta$ ratio, and spectral index volume: (1) sources with steep radio spectral index, large radio loudness and large [OIII]/H$\beta$ ratio; (2) sources with flat radio spectral index, intermediate radio loudness and lower [OIII]/H$\beta$ ratio; (3) sources with inverted radio spectral index, low radio loudness and low [OIII]/H$\beta$. The groups (1), (2), and (3) are located along the Seyfert-LINER spectral classes towards lower ionization ratios in the ODDs and hence can represent different activity cycles/accretion modes of AGNs, which could be in some cases associated with different merger stages.

\end{abstract}

\section{Introduction}

\begin{figure}[tbh]
  \centering
  \includegraphics[width=0.7\textwidth]{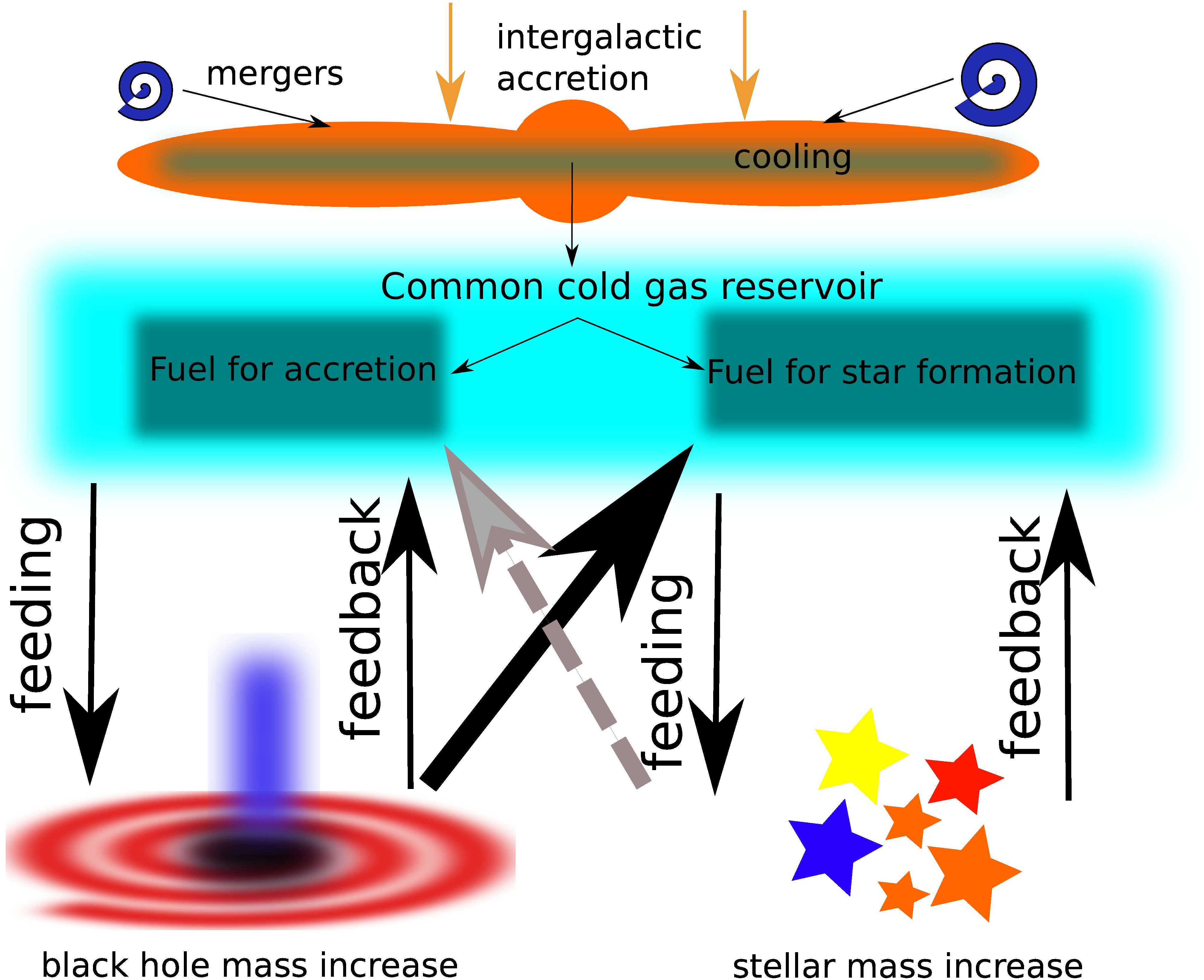} 
  \caption{Illustration of the feeding and the feedback process of an AGN within its galactic host. From the top part, the figure illustrates the source of a gaseous material (intergalactic accretion, gas-rich mergers, internal galactic sources). Cold gas formed via the radiative and the adiabative cooling fuels both the black hole accretion as well as the star formation. Both processes provide energetic and mechanical feedback to the cold gas reservoir. Inspired by \citet{2017NatAs...1E.165H}.}
  \label{fig_feedback}
\end{figure}

There are several correlations between the mass of supermassive black holes and the properties of host bulges, namely the stellar velocity dispersion, its mass, and its luminosity \citep{2013ARA&A..51..511K}. These correlations point towards a common evolution of the central massive black hole and the surrounding host. So far the general understanding of the galaxy evolution relies on the negative feedback of active phases on the star-formation rate in the host. The enhanced accretion activity, so-called active galactic nucleus (hereafter AGN), is a source of X-ray/UV radiation and the kinetic energy that leads to heating of the cold molecular gas and/or increasing its turbulence. The imparted momentum can also blow some of the cold gas away. All of these processes generally lead to the lack of cold molecular gas, from which stars can form and hence the overall star-formation is significantly reduced \citep{2017NatAs...1E.165H}. On the other hand, the momentum of the nuclear outflow/jet can also compress the cold gas, which leads to a star-formation episode, i.e. a positive AGN feedback \citep{2018ApJ...852...63M}. Although the overall picture is quite complex, see the illustration in Fig.~\ref{fig_feedback}, the accretion and star-formation are fueled from the same gas reservoir and hence these processes are generally coupled to each other with a certain time-delay \citep{2006NewAR..50..677H}. The removal of cold gas is expected to eventually lead to a low star-formation rate and the low accretion rate, which is the stage of quiescent galaxies and low-luminosity AGN, of which the Milky Way with Sgr A* compact source is a prototype \citep{2017FoPh...47..553E}.

The role of radio-loud galaxies in the AGN feedback and the overall galaxy evolution is currently unclear. Therefore, it is vital to study the relations between radio and optical properties of galaxies to better comprehend the role of nuclear and jet activity on the large-scale galaxy evolution. To this end, we performed radio observations between 4.85 and 10.45 GHz to determine radio spectral slopes of intermediate-redshift galaxies. This information was complemented by radio-loudness and optical narrow-line emission ratios.

This contribution is structured as follows. In Section~\ref{sec_SDSS-FIRST}, we introduce the sample of SDSS-FIRST sources. In Section~\ref{sec_spectral_index_spectral_classes}, we present the basic characteristics of radio spectral index distributions in different optical spectral classes of galaxies. Subsequently, the distribution of radio spectral indices in optical diagnostic diagrams is presented in Section~\ref{sec_spectral_index_ODD}. Finally, we conclude with Section~\ref{sec_conclusions}.   

\section{SDSS-FIRST sample}
\label{sec_SDSS-FIRST}

To study the correlations between radio and optical properties across the optical diagnostic diagram of galaxies, we selected 119 intermediate-redshift radio-optical galaxies ($0.04\lesssim z \lesssim 0.4$) in the cross-matched SDSS-FIRST surveys (Sloan Digital Sky Survey - SDSS, \citealt{2000AJ....120.1579Y}; Faint Images of the Radio Sky at Twenty Centimeters-FIRST, \citealt{1995ApJ...450..559B}) with 1.4 GHz flux densities in the range of $100\,{\rm mJy}<F_{1.4}<1000\,{\rm mJy}$ and 396 sources with lower flux densities in the range $10\,{\rm mJy}<F_{1.4}<100\,{\rm mJy}$, see \citet{2012A&A...546A..17V}, \citet{2015A&A...573A..93V}, and \citet{2019A&A...630A..83Z} for details. Our cross-matched sample with the lower radio flux limit of 10 mJy  has two biases:
\begin{itemize}
  \item in comparison with the parent sample of SDSS-FIRST galaxies, which are predominantly located in the star-forming/composite region of the Baldwin, Phillips, and Terlevich (BPT) ODD, our sample is shifted towards higher [OIII]/H$\beta$ ratio, that is in the composite-Seyfert-LINER\footnote{LINER stands for a low-ionization nuclear emission-line region.} region,
  \item because of the detectability of the sources at the higher radio frequencies at 4.85 and 10.45 GHz (the sensitivity limit of the Effelsberg radio telescope is $\sim 5\,{\rm mJy}$), our sample is biased towards flat-spectrum sources.
\end{itemize} 
The Effelsberg observations at 4.85 GHz and 10.45 GHz of the sources are in the catalogues by \citet{2015A&A...573A..93V} and \citet{2019A&A...630A..83Z}.

\section{Radio spectral index distribution in different optical spectral classes}
\label{sec_spectral_index_spectral_classes}

Using different spectral classes of galaxies in the BPT diagram, namely star-forming, composite, AGN Seyfert, and LINER sources, we first looked at the fractions of steep, flat, and inverted sources in each spectral class. We performed this analysis for both lower frequencies (1.4 GHz-4.85 GHz, $\alpha_{[1.4-4.85]}$) and higher frequencies (4.85 GHz-10.45 GHz, $\alpha_{[4.85-10.45]}$). The mean, the median, $16\%$-, and $84\%$-percentile of the radio spectral index for lower and higher frequencies for each optical spectral class are listed in Table~\ref{tab_spectralindex_1.4_4.85} and Table~\ref{tab_spectralindex_4.85_10.45}, respectively. As we showed in \citet{2019A&A...630A..83Z}, LINER sources contain the larger fraction of flat and inverted sources than Seyfert galaxies at both lower (82.4\% vs. 71.8\%) and higher frequencies (66.3\% vs. 58.3\%). As it can also be inferred from Table~\ref{tab_spectralindex_1.4_4.85} and Table~\ref{tab_spectralindex_4.85_10.45}, LINERs have larger both the mean and the median spectral index in comparison with Seyfert sources, which implies the presence of compact self-absorbed radio cores and core-jet systems in the majority of these sources. Hence, in our sample, AGN contributes to the ionizing continuum in LINERs. In other words, LINERs in our sample are not characterized by a missing AGN, which was advocated previously \citep{2013A&A...558A..43S}.

\begin{table}[h!]
  \centering
   \caption{Radio spectral index $\alpha_{[1.4-4.85]}$ distribution among different optical spectral classes. From the left to the right: mean, standard deviation, median, $16\%$-, and $84\%$-percentile of the radio spectral index $\alpha_{[1.4-4.85]}$, respectively, for each optical spectral class of galaxies and the whole sample.}
  \resizebox{\linewidth}{!}{
  \begin{tabular}{cccccc}
    \hline
    \hline
    Spectral class & Mean $\alpha_{[1.4-4.85]}$ & $\sigma$ & Median $\alpha_{[1.4-4.85]}$ & $16\%\,P$ & $84\%\,P$\\
    \hline
    Star-forming & $-0.25$ & $0.45$ & $-0.33$ & $-0.66$ & $0.24$\\
    Composite & $-0.25$ & $0.54$ & $-0.40$ & $-0.73$ & $0.22$\\
    Seyfert & $-0.31$ & $0.61$ & $-0.49$ & $-0.83$ & $0.21$\\
    LINER & $-0.22$ & $0.50$ & $-0.24$ & $-0.72$ & $0.26$\\
    \hline 
    Total & $-0.25$ & $0.54$ & $-0.36$ & $-0.76$ & $0.24$\\
    \hline    
  \end{tabular}} 
  \label{tab_spectralindex_1.4_4.85}
\end{table}

\begin{table}[h!]
  \centering
   \caption{Radio spectral index $\alpha_{[4.85-10.45]}$ distribution among different optical spectral classes. From the left to the right: mean, standard deviation, median, $16\%$-, and $84\%$-percentile of the spectral index $\alpha_{[4.85-10.45]}$, respectively, for each optical spectral class of galaxies and the whole sample.}
   \resizebox{\linewidth}{!}{
  \begin{tabular}{cccccc}
    \hline
    \hline
    Spectral class & Mean $\alpha_{[4.85-10.45]}$ &  $\sigma$ & Median $\alpha_{[4.85-10.45]}$ & $16\%\,P$ & $84\%\,P$\\
    \hline
    Star-forming & $-0.61$ &  $0.65$ & $-0.65$ & $-1.02$ & $0.05$\\
    Composite & $-0.42$ & $0.80$ & $-0.43$ & $-0.90$ & $0.19$\\
    Seyfert & $-0.63$ & $0.52$ & $-0.64$ & $-0.89$ & $-0.29$\\
    LINER & $-0.46$ & $0.59$ & $-0.59$ & $-0.95$ & $0.02$\\
    \hline
    Total & $-0.51$ & $0.63$ & $-0.58$ & $-0.92$ & $0.00$\\
    \hline    
  \end{tabular}} 
  \label{tab_spectralindex_4.85_10.45}
\end{table}

\section{Distribution of radio spectral indices of galaxies in narrow-line optical diagnostic diagrams} 
\label{sec_spectral_index_ODD}

\begin{figure}[h!]
    \includegraphics[width=\textwidth]{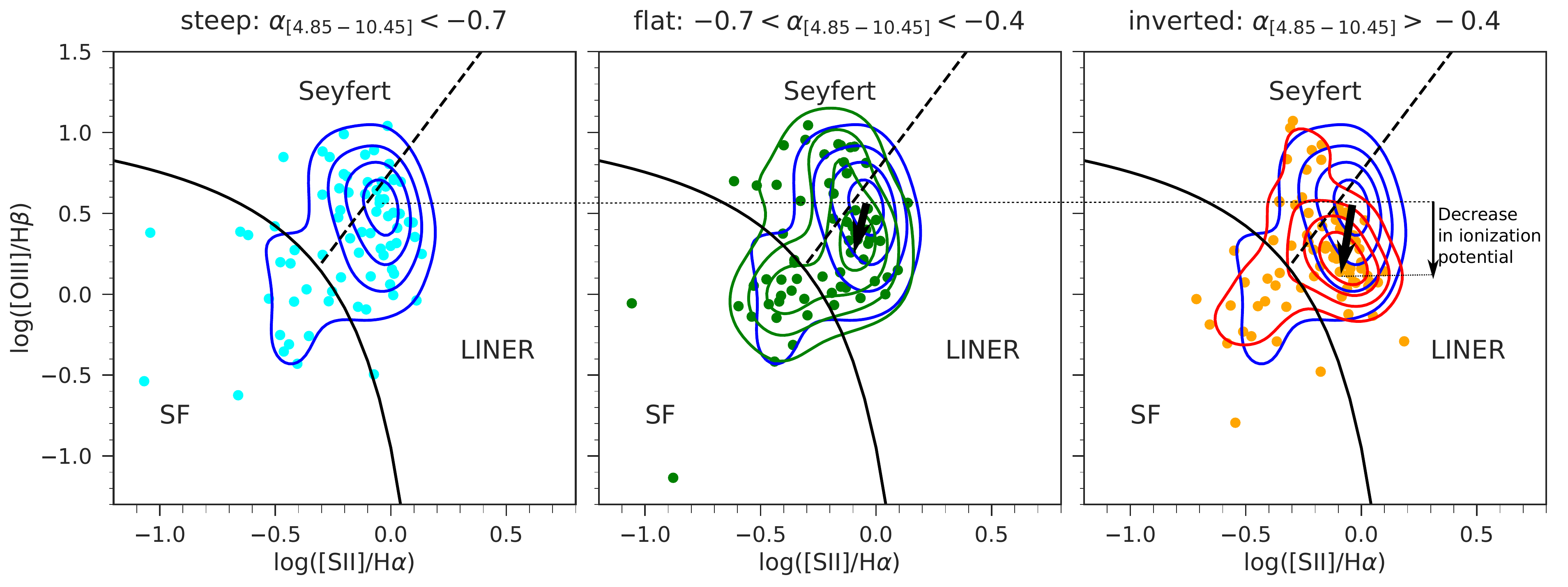}
    \caption{Distribution of steep, flat, and inverted spectral indices in [SII]/H$\alpha$-based optical diagnostic diagram. From the left to the right panel, we show the distribution of the steep, the flat, and the inverted spectral index. The right panel depicts the decrease in the line ratio $\log$([OIII]/H$\beta$) between the steep and the inverted sources. The abbreviation SF stands for star-forming galaxies.}
    \label{fig_BPT_SII_specindex}
\end{figure}

Now we will consider how different radio spectral indices (steep, flat, and inverted) are distributed in the optical diagnostic diagrams. The first visible effect is the gradual vertical shift of the sources with the larger spectral index (flat to inverted) towards a lower ionization potential represented by the narrow line ratio $\log$([OIII]/H$\beta$). In other words, steep sources are concentrated towards the Seyfert AGN spectral part characterized by the larger $\log$([OIII]/H$\beta$), while inverted sources are positioned more towards the LINER part with the lower ratio $\log$([OIII]/H$\beta$), see Fig.~\ref{fig_BPT_SII_specindex}.

To gain a better insight into the distribution of radio spectral indices with respect to the optical and radio properties, we calculated the radio loudness for each source using the following definition,
\begin{equation}
R_{\rm g}\equiv \log{\left(\frac{F_{\rm radio}}{F_{\rm optical}}\right)}=0.4(g-m_{1.4})\,,
\label{eq_radio_loudness}
\end{equation}
where $g$ is the magnitude corresponding to the optical $g$-band and $m_{1.4}$ is the magnitude corresponding to the 1.4 GHz flux density. 

As a next step, we applied the unsupervised machine-learning Gaussian Mixture Model (GMM) to our data. We looked for the clusterings in the 3D space of the radio-loudness $R_{\rm g}$, narrow-line ratio $\log$([OIII]/H$\beta$), and the radio spectral slope $\alpha_{[4.85-10.45]}$, i.e., each source is represented by the vector,

\begin{equation}
\vec{x} = (R_g,\log([\mathrm{OIII}]/\mathrm{H}\beta),\alpha_{[4.85-10.45]}).
\end{equation}

The GMM works with the probability density function given by the weighted sum of individual Gaussian components with unknown parameters. The output of the GMM is the probability of each source belonging to each of the components. In our model, we assumed three components motivated by three radio spectral classes (steep, flat, and inverted) and we made use of the expectation maximization technique.  

\begin{figure}
\begin{center}
\includegraphics[width=0.8\textwidth]{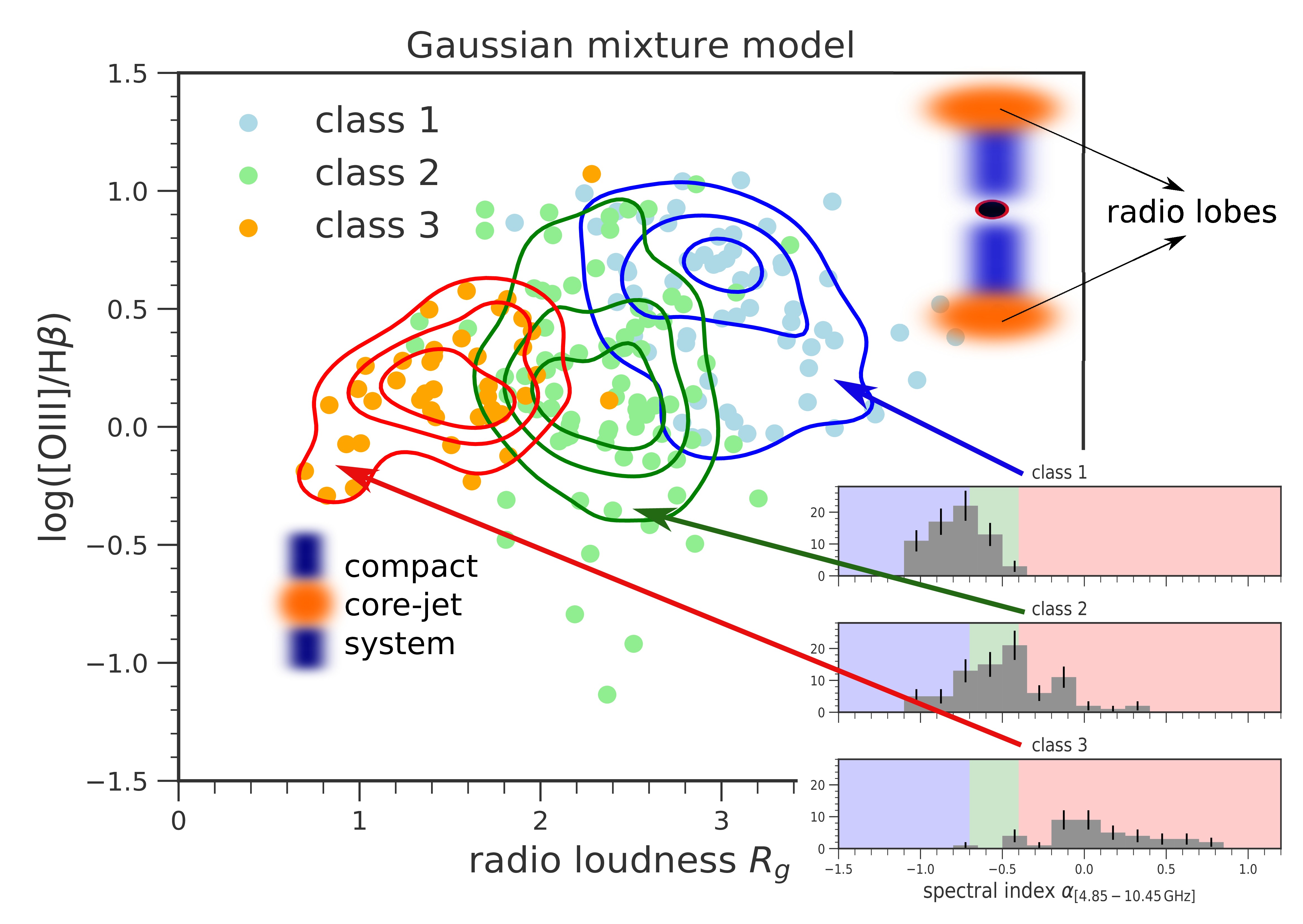}
\end{center}
  \caption{Three classes of radio-optical emitters in the radio-loudness, ionization ratio, and radio spectral index space. The histograms of radio spectral indices for the three classes are displayed in the right plot inset. The illustration in the upper right corner represents older radio lobes as the source of optically thin synchrotron emission, while the illustration in the lower-left corner stands for a compact radio core-jet system as the source of inverted-flat self-absorbed, optically thick synchrotron emission.}
  \label{fig_GMM}
\end{figure}

We obtain three basic clusters in the $R_{\rm g}-\log$([OIII]/H$\beta$) plane, see Fig.~\ref{fig_GMM}, where the third dimension is represented by a colour-coded spectral index. The corresponding distributions of the radio spectral index are displayed in the right panel inset. These three clusters represent three classes of sources:
\begin{itemize}
  \item[(1)] sources with a steep radio spectral index ($\alpha_{[4.85-10.45]}<-0.7$) with a high ionization ratio and a large radio loudness,
  \item[(2)] sources with a flat radio spectral index ($-0.7<\alpha_{[4.85-10.45]}<-0.4$) with a lower ionization ratio and an intermediate radio loudness,
  \item[(3)] sources with an inverted radio spectral index ($\alpha_{[4.85-10.45]}>-0.4$) with a low ionization ratio and a small radio loudness.
\end{itemize} 
Since our sources have intermediate redshift after the peak of quasar/star-formation activity, classes (1), (2), and (3) can represent different stages of the intermittent accretion activity, which for radio galaxies evolves with the typical timescale of $\sim 10^3-10^4$ years due to the radiation pressure instability within an accretion disc \citep{2009ApJ...698..840C}. Apart from accretion-disc instabilities, mergers can also trigger and influence the jet activity and its dynamics, see e.g., detailed studies of OJ287 and 3C84 \citep{2018MNRAS.478.3199B,2019Galax...7...72B}.

\section{Conclusions}
\label{sec_conclusions}

For the sample of intermediate-redshift galaxies, we studied the distribution of their radio spectral indices between 4.85 and 10.45 GHz in the optical diagnostic diagrams. Using the machine learning Gaussian Mixture Model, we found that steep, flat, and inverted sources occupy different regions of the radio-loudness--ionization ratio plane, with the steep sources located towards a larger radio-loudness as well as a higher ionization ratio, while the flat/inverted sources are situated in the region of a smaller radio-loudness and a lower ionization ratio. In terms of optical spectral classes of galaxies, steep sources are located towards the Seyfert region, which indicates the presence of large-scale jets and outflows for these sources. In comparison, the flat/inverted sources occupy the LINER region, which implies the presence of a compact radio-core-jet system for LINERs.

\acknowledgements{MZ and BC acknowledge the financial support by the National Science Centre, Poland, grant No. 2017/26/A/ST9/00756 (Maestro 9). Previously, the project has been supported by SFB956 ``Conditions and impact of star formation: Astrophysics, Instrumentation, and Laboratory Research'', where MZ, AE, and SB have been members of the subgroup A2 ``Conditions for star formation in nearby AGN and QSO hosts''.}

%\bibliographystyle{ptapap}
%\bibliography{pta_zajacek}

\end{document}